\documentclass[12pt]{article}

\begin{document}

\begin{center}
{\Large\bf{}Quasilocal energy-momentum for tensors B and V in small
regions}
\end{center}

\begin{center}
Lau Loi So\\
Department of Physics, National Central University, Chung-Li 320, Taiwan\\
Department of Physics, Tamkang University, Tamsui 251, Taiwan
\end{center}



\begin{abstract}
The Bel-Robinson tensor $B$ and the tensor $V$ have the same
quasilocal energy-momentum in a small sphere. Using a pseudotensor
approach to evaluate the energy-momentum in a half-cylinder, we
find that $B$ and $V$ have different values, not proportional to
the ``Bel-Robinson energy-momentum". Furthermore, even if we
arrange things so that we do get the same ``Bel-Robinson
energy-momentum" value, the angular momentum gives different
values using $B$ and $V$ in a half cylinder. In addition, we find
that $B$ and $V$ have a different number of independent
components. The fully trace free property of $B$ and $V$ implies
conservation of pure ``Bel-Robinson energy-momentum" in small
regions, and vice versa. In addition, we also demonstrate the
tidal heating, rate of change of momentum and spin angular momentum flux by using these two
tensors.
\end{abstract}

\section{Introduction}
In attempts to identify a good physical expression for the local
distribution of gravitational energy-momentum there have been many
different approaches which are similar to
Einstein's~\cite{Trautman}. For example, those of
Landau-Lifshitz~\cite{LL}, Bergmann-Thomson~\cite{Bergmann},
Papapetrou~\cite{Papapetrou} and Weinberg~\cite{Weinberg}. Most of
them deal with the Einstein equation:
$G_{\mu\nu}=\kappa{}T_{\mu\nu}$, where $\kappa$ is a constant,
$G_{\mu\nu}$ and $T_{\mu\nu}$ are the Einstein and stress tensors.
One can define a superpotential with a suitable anti-symmetry
$U_{\alpha}{}^{\mu\nu}\equiv{}U_{\alpha}{}^{[\mu\nu]}$ and remove
a divergence of $U_{\alpha}{}^{\mu\nu}$ from $G_{\mu\nu}$ to
define the gravitational energy-momentum density
\begin{equation}
2\kappa{}\mathbf{t}_{\alpha}{}^{\mu}
:=\partial_{\nu}U_{\alpha}{}^{[\mu\nu]}
-2\sqrt{-g}\,G_{\alpha}{}^{\mu}.
\end{equation}
Note that $\mathbf{t}_{\alpha}{}^{\mu}$ is a
pseudotensor~\cite{SoNesterPRD2009}. Using the Einstein equation,
we have a total energy-momentum density which satisfies
\begin{equation}
\partial_{\nu}U_{\alpha}{}^{[\mu\nu]}=2\kappa{\cal{}T}_{\alpha}{}^{\mu}
=2\kappa\sqrt{-g}(\mathbf{T}_{\alpha}{}^{\mu}+\mathbf{t}_{\alpha}{}^{\mu}),\label{24bDec2021}
\end{equation}
where $\mathbf{T}_{\alpha}{}^{\mu}=\sqrt{-g}\,T_{\alpha}{}^{\mu}$
and hence, due to antisymmetry of $U_{\alpha}{}^{[\mu\nu]}$, is
automatically conserved, i.e., has a vanishing divergence.

The proposed criteria for testing quasilocal expressions included:
(i) limit to good weak field values (i.e., linearized gravity).
(ii) good asymptotic values both at spatial and null infinity. We
here emphasize that the criteria for these two are not very
restrictive; they only test the quasilocal expression to linear
order. (iii) positivity (i.e., globally) is a strong test but is
not easy to achieve, (iv) small region inside of matter: the
quasilocal energy-momentum expression should, by the equivalence
principle, reduce to the material source terms. Most classical
pseudotensors pass this test.  (v) small region vacuum: positivity
for the first non-vanishing parts of the quasilocal expression.
This depends on the gravitational field non-linearly, and hence it
can give a discriminating test of the expression, it is quite
non-trivial but not impossibly difficult.

Positive quasilocal gravitational energy should hold not only on a
large scale but also on the small scale~\cite{Szabados}. However
it is generally not at all easy to prove that a particular
expression enjoys this property. A good test case is the small
region limit. This will be our concern in this work.  Here we
consider specifically the pseudotensor expressions.  For a small
region, one can expand the energy-momentum density in Riemann
normal coordinates (RNC) about the origin:
\begin{eqnarray}
{\cal{}T}_{\alpha}{}^{\beta}(x)
&=&{\cal{T}}_{\alpha}{}^{\beta}|_{0}+\partial_{\mu}{\cal{}T}_{\alpha}{}^{\beta}|_{0}x^{\mu}
+\frac{1}{2}\partial^{2}_{\mu\nu}{\cal{T}}_{\alpha}{}^{\beta}|_{0}x^{\mu}x^{\nu}+...\nonumber\\
&=&\mathbf{T}_{\alpha}{}^{\beta}|_{0}+\partial_{\mu}\mathbf{T}_{\alpha}{}^{\beta}|_{0}\,x^{\mu}+...
+\mathbf{t}_{\alpha}{}^{\beta}|_{0}+\partial_{\mu}\mathbf{t}_{\alpha}{}^{\beta}|_{0}\,x^{\mu}
+\frac{1}{2}\partial^{2}_{\mu\nu}\mathbf{t}_{\alpha}{}^{\beta}|_{0}x^{\mu}x^{\nu}+....
\end{eqnarray}
By construction $\mathbf{t}_{\alpha}{}^{\beta}|_{0}$ and
$\partial_{\mu}\mathbf{t}_{\alpha}{}^{\beta}|_{0}$ vanish in
vacuum. Consequently, for small $x^{\mu}$ inside of matter the
$\mathbf{T}_{\alpha}{}^{\beta}$ and
$\partial_{\mu}\mathbf{T}_{\alpha}{}^{\beta}$ terms dominate (this
is a reflection of the equivalence principle). In vacuum regions
all the $\mathbf{T}_{\alpha}{}^{\beta}$ terms vanish, then the
lowest order non-vanishing term is
$\frac{1}{2}\partial^{2}_{\mu\nu}\mathbf{t}_{\alpha}{}^{\beta}|_{0}x^{\mu}x^{\nu}$.
This is the object on which we focus our attention in this work.
It turns out that for all proposed pseudotensor and quasilocal
energy-momentum expressions this fourth rank tensor is quadratic
in the Riemann (equivalent in empty space regions to the Weyl)
tensor. That is why the quadratic curvature expressions become
interesting and important (i.e.,
$\partial^{2}_{\mu\nu}\mathbf{t}_{\alpha}{}^{\beta}\simeq{}
R_{....}R_{....}$). Normally, the expansion of a pseudotensor
expression up to second order can only be some linear combination
of three tensors $\{B,S,K\}$ or
$\{B,V,S\}$~\cite{SoNesterPRD2009,MTW,SoCQG2009} which are each
certain quadratic expressions in the curvature.

According to a review article~(4.2.2 in~\cite{Szabados}):
``Therefore, in vacuum in the leading $r^{5}$ order any coordinate
and Lorentz-covariant quasilocal energy-momentum expression which
is non-spacelike and future pointing must be proportional to the
Bel-Robinson `momentum':
$B_{\mu\lambda\xi\kappa}t^{\lambda}t^{\xi}t^{\kappa}$." Note that
here $t^{\alpha}$ is timelike unit vector and `momentum' means
4-momentum (see (28)).  This is a strong test. The Bel-Robinson
tensor $B$ has many nice properties such as fully symmetric,
traceless and divergence free~\cite{Senovilla}. It is known that
$B$ contributes positivity in a small sphere region and perhaps it
maybe thought that it is the only one. However, we recently
proposed an alternative $V$ (see (18)) which has the identical
``Bel-Robinson momentum" at the same limit, i.e.,
$(B_{\mu\lambda\xi\kappa}-V_{\mu\lambda\xi\kappa})t^{\lambda}t^{\xi}t^{\kappa}\equiv0$.
Confined to a small spherical or cubical regions~\cite{Garecki},
$B$ and $V$ cannot be distinguished. One may suspect that $V$ is
redundant because $B$ can manage all the jobs, But we claim not.

As the basic requirement for the quasilocal energy is any closed
2-surface, we examined the energy-momentum and angular momentum in
other regions (see Table 1). we find for the energy in a small
half-cylinder when $h\neq\sqrt{3}a$ give different values if
substituting $\mathbf{t}$ by $B$ and $V$, which means that they
are distinguishable. Only for one particular ratio $h=\sqrt{3}a$,
$B$ and $V$ both give the same ``Bel-Robinson momentum" value,
however we lose the distinction between them again. Therefore we
turn to examining the angular momentum in a small half-cylinder,
and show that when we replacing $\mathbf{t}$ by $B$ and $V$ in the
angular momentum expression they contribute different values,
thereby clarifying that the two tensors  are really
distinguishable.

Here we remark some components of the angular momentum in a
hemi-sphere show that $B$ contributes a null result while $V$
gives non-zero values (see section 3.2). The reason comes from the
fully symmetric property of $B$, while $V$ only has some certain
symmetry property (see (19)). Consequently, $V$ is
non-replaceable.

Moreover, we also demonstrate the tidal heating, rate of change of momentum and spin angular
momentum flux by using tensors $B$ and $V$. Once again, they have
the same results.

\section{Technical background}
Using a Taylor series expansion, the metric tensor can be written
as
\begin{eqnarray}
g_{\alpha\beta}(x^{\lambda}) =g_{\alpha\beta}|_{x^{\lambda}_{0}}
+\partial_{\mu}g_{\alpha\beta}|_{x^{\lambda}_{0}}(x^{\mu}-x^{\mu}_{0})
+\frac{1}{2}\partial^{2}_{\mu\nu}g_{\alpha\beta}|_{x^{\lambda}_{0}}
(x^{\mu}-x^{\mu}_{0})(x^{\nu}-x^{\nu}_{0})+...,
\end{eqnarray}
where the metric signature is +2. For simplicity, let
$x^{\lambda}_{0}=0$ and at the origin in RNC
\begin{eqnarray}
g_{\alpha\beta}|_{0}&=&\eta_{\alpha\beta},\quad\quad
\partial_{\mu}g_{\alpha\beta}|_{0}=0,\\
-3\partial^{2}_{\mu\nu}g_{\alpha\beta}|_{0}
&=&R_{\alpha\mu\beta\nu}+R_{\alpha\nu\beta\mu},\quad\quad
-3\partial_{\nu}\Gamma^{\mu}{}_{\alpha\beta}|_{0}
=R^{\mu}{}_{\alpha\beta\nu}+R^{\mu}{}_{\beta\alpha\nu}.
\end{eqnarray}
Three basic tensors~\cite{SoNesterPRD2009,MTW,SoCQG2009} that
commonly occurred in the pseudotensor are:
\begin{eqnarray}
&&B_{\alpha\beta\mu\nu}\equiv{}B_{(\alpha\beta\mu\nu)}
:=R_{\alpha\lambda\mu\sigma}R_{\beta}{}^{\lambda}{}_{\nu}{}^{\sigma}
+R_{\alpha\lambda\nu\sigma}R_{\beta}{}^{\lambda}{}_{\mu}{}^{\sigma}
-\frac{1}{8}g_{\alpha\beta}g_{\mu\nu}R^{2}_{\lambda\sigma\rho\tau},\label{24aDec2021}\\
&&S_{\alpha\beta\mu\nu}\equiv{}S_{(\alpha\beta)(\mu\nu)}\equiv{}S_{\mu\nu\alpha\beta}
:=R_{\alpha\mu\lambda\sigma}R_{\beta\nu}{}^{\lambda\sigma}
+R_{\alpha\nu\lambda\sigma}R_{\beta\mu}{}^{\lambda\sigma}
+\frac{1}{4}g_{\alpha\beta}g_{\mu\nu}R^{2}_{\lambda\sigma\rho\tau},\label{29April2009}\\
&&K_{\alpha\beta\mu\nu}\equiv{}K_{(\alpha\beta)(\mu\nu)}\equiv{}K_{\mu\nu\alpha\beta}
:=R_{\alpha\lambda\beta\sigma}R_{\mu}{}^{\lambda}{}_{\nu}{}^{\sigma}
+R_{\alpha\lambda\beta\sigma}R_{\nu}{}^{\lambda}{}_{\mu}{}^{\sigma}
-\frac{3}{8}g_{\alpha\beta}g_{\mu\nu}R^{2}_{\lambda\sigma\rho\tau},
\end{eqnarray}
where
$R^{2}_{\lambda\sigma\rho\tau}=R^{\lambda\sigma\rho\tau}R_{\lambda\sigma\rho\tau}$.


It may be worthwhile to emphasize that $B$ has a very good analog
with the electromagnetic energy-momentum tensor
$\mathbf{T}^{\mu\nu}$. In Minkowski coordinates $(t,x,y,z)$:
\begin{eqnarray}
\mathbf{T}^{00}&=&\frac{1}{2}\left(E_{a}E^{a}+B_{a}B^{a}\right),\\
\mathbf{T}^{0i}&=&\delta^{ij}\epsilon_{jab}E^{a}B^{b}~=~(\vec{E}\times\vec{B})^{i},\\
\mathbf{T}^{ij}&=&\frac{1}{2}\left[\delta^{ij}\left(E_{a}E^{a}+B_{a}B^{a}\right)
-2\left(E^{i}E^{j}+B^{i}B^{j}\right)\right].
\end{eqnarray}
where $\vec{E}$ and $\vec{B}$ refer to the electric and magnetic
field density. In order to appreciate the nice properties of $B$,
we compare the energy density with $S$ and $K$
\begin{eqnarray}
B_{0000}=E^{2}_{ab}+H^{2}_{ab},\quad{}S_{0000}=2(E^{2}_{ab}-H^{2}_{ab}),
\quad{}K_{0000}=-E^{2}_{ab}+3H^{2}_{ab}.\label{3aDec2011}
\end{eqnarray}
where the evaluation has used the electric part $E_{ab}$ and
magnetic part $H_{ab}$, defined in terms of the Weyl
tensor~\cite{Carmeli}: $E_{ab}:=C_{a0b0}$ and $H_{ab}:=*C_{a0b0}$
where $*C_{\alpha\beta\mu\nu}$ means its dual. Likewise for the
momentum density (i.e., Poynting vector)
\begin{equation}
B_{000i}=2\epsilon_{ij\,k}E^{j\,d}H^{k}{}_{d},\quad{}
S_{000i}=0,\quad{}
K_{000i}=2\epsilon_{ij\,k}E^{j\,d}H^{k}{}_{d}.\label{3bDec2011}
\end{equation}
Finally, the stress,
\begin{eqnarray}
B_{00ij}&=&\delta_{ij}(E_{ab}E^{ab}+H_{ab}H^{ab})-2(E_{id}E_{j}{}^{d}+H_{id}H_{j}{}^{d}),\label{3cDec2011}\\
S_{00ij}&=&-2\left[\delta_{ij}(E_{ab}E^{ab}-H_{ab}H^{ab})+2(E_{id}E_{j}{}^{d}-H_{id}H_{j}{}^{d})\right],\\
K_{00ij}&=&\delta_{ij}(5E_{ab}E^{ab}-3H_{ab}H^{ab})-4E_{id}E_{j}{}^{d}.\label{3dDec2011}
\end{eqnarray}
We observe that summing up $S$ and $K$ has exactly the same energy
as $B$: $(B_{0000}-S_{0000}-K_{0000})
\equiv{}0\equiv(B_{00ij}-S_{00ij}-K_{00ij})\delta^{ij}$. It is
natural to define the alternative 4th rank tensor~\cite{SoCQG2009}
as follows
\begin{equation}
V:=S+K\equiv{}B+W_,\label{24Feb2009}
\end{equation}
where
\begin{equation}
W_{\alpha\beta\mu\nu}:=\frac{3}{2}S_{\alpha\beta\mu\nu}
-\frac{1}{8}(5g_{\alpha\beta}g_{\mu\nu}
-g_{\alpha\mu}g_{\beta\nu}-g_{\alpha\nu}g_{\beta\mu})R^{2}_{\lambda\sigma\rho\tau}.\label{3aMar2010}
\end{equation}
This is the tensor that we prefer to focus on. Both $V$ and $W$
satisfy the following properties
\begin{eqnarray}
&&X_{\alpha\beta\mu\nu}\equiv{}X_{(\alpha\beta)(\mu\nu)}\equiv{}X_{\mu\nu\alpha\beta},\quad{}
X_{\alpha\beta\mu}{}^{\mu}\equiv{}0\equiv{}X_{\alpha\mu\beta}{}^{\mu}.\label{3eDec2011}
\end{eqnarray}
It is known $B$ has the dominant energy
property~\cite{Senovilla,Penrose}:
$B_{\alpha\beta\mu\nu}w_{1}^{\alpha}w_{2}^{\beta}w_{3}^{\mu}w_{4}^{\nu}\geq{}0$,
where $w_{1},w_{2},w_{3},w_{4}$ are any future-pointing causal
vectors. Intuitively, referring to (\ref{24Feb2009}), $V$ may
contain more non-trivial independent components than $B$ and
indeed it is the case (see section (3.3)). While $V$ only
satisfies the weak energy condition and $W$ fulfills none of them.
For $W$, it does not contribute energy-momentum in small sphere.
For a comparison of $B$ and $V$, we find that it is more
convenient to use $(B+W)$ instead of $(S+K)$ for the
representation of $V$.

In our work, we are mainly dealing with expression of the 4th rank
which are quadratic in the curvature tensor. There are four
tensors which form a basis with appropriate
symmetries~\cite{Deser}, we use
\begin{eqnarray}
&&\tilde{B}_{\alpha\beta\mu\nu}
:=R_{\alpha\lambda\mu\sigma}R_{\beta}{}^{\lambda}{}_{\nu}{}^{\sigma}
+R_{\alpha\lambda\nu\sigma}R_{\beta}{}^{\lambda}{}_{\mu}{}^{\sigma}
,\quad{}\tilde{S}_{\alpha\beta\mu\nu}
:=R_{\alpha\mu\lambda\sigma}R_{\beta\nu}{}^{\lambda\sigma}
+R_{\alpha\nu\lambda\sigma}R_{\beta\mu}{}^{\lambda\sigma}
,\label{29aApril2009}\\
&&\tilde{K}_{\alpha\beta\mu\nu}
:=R_{\alpha\lambda\beta\sigma}R_{\mu}{}^{\lambda}{}_{\nu}{}^{\sigma}
+R_{\alpha\lambda\beta\sigma}R_{\nu}{}^{\lambda}{}_{\mu}{}^{\sigma},\quad{}
\tilde{T}_{\alpha\beta\mu\nu}
:=-\frac{1}{8}g_{\alpha\beta}g_{\mu\nu}R^{2}_{\lambda\sigma\rho\tau}.\label{20bJan2009}
\end{eqnarray}
They are designed to describe the gravitational energy expression
based on the pseudotensor (see (\ref{25Sep2008})) and are
manifestly symmetric in the last two indices, i.e.,
$\tilde{M}_{\alpha\beta\mu\nu}=\tilde{M}_{\alpha\beta(\mu\nu)}$.
Then
$\tilde{M}_{\alpha\beta\mu\nu}=\tilde{M}_{(\alpha\beta)\mu\nu}$
and it also naturally turns out that
$\tilde{M}_{\alpha\beta\mu\nu}=\tilde{M}_{\mu\nu\alpha\beta}$. In
addition, here we write down another form of the representation
for this Bel-Robinson tensor:
\begin{equation}
B_{\alpha\beta\mu\nu}\equiv-\frac{1}{2}S_{\alpha\beta\mu\nu}
+K_{\alpha\beta\mu\nu}+\frac{1}{8}(5g_{\alpha\beta}g_{\mu\nu}
-g_{\alpha\mu}g_{\beta\nu}-g_{\alpha\nu}g_{\beta\mu})R^{2}_{\lambda\sigma\rho\tau}.
\label{29Oct2008}
\end{equation}

Here come to the situation for applying the tidal heating, rate of
change of momentum and time dependent spin angular momentum by
using the tensors $B$ and $V$. In weak field the metric tensor can
be decomposed as
$g_{\alpha\beta}=\eta_{\alpha\beta}+h_{\alpha\beta}$, and its
inverse $g^{\alpha\beta}=\eta^{\alpha\beta}-h^{\alpha\beta}$. Here
we mainly use the first order and ignore the higher orders.
According to Zhang~\cite{Zhang}, the metric components can be
written as
\begin{eqnarray}
h^{00}&=&\frac{3}{r^{5}}I_{ab}x^{a}x^{b}-E_{ab}x^{a}x^{b},\\
h^{0j}&=&\frac{4}{r^{5}}\,\epsilon^{j}{}_{pq}J^{p}{}_{l}\,x^{q}x^{l}
+\frac{2}{3}\epsilon^{j}{}_{pq}B^{p}{}_{l}x^{q}x^{l}
+\frac{2}{r^{3}}\dot{I}^{j}{}_{a}\,x^{a}
+\frac{10}{21}\dot{E}_{ab}x^{a}x^{b}x^{j}
-\frac{4}{21}\dot{E}^{j}{}_{a}x^{a}r^{2},\\
h^{ij}&=&\eta^{ij}h^{00}+\tilde{h}^{ij},
\end{eqnarray}
where
\begin{eqnarray}
\tilde{h}^{ij}=\frac{8}{3r^{3}}\epsilon_{pq}{}^{(i}\dot{J}^{j)p}x^{q}
+\frac{5}{21}x^{(i}\epsilon^{j)}{}_{pq}\dot{B}^{q}{}_{l}x^{p}x^{l}
-\frac{1}{21}r^{2}\epsilon_{pq}{}^{(i}\dot{B}^{j)\,q}x^{p}.
\end{eqnarray}
Zhang used $\tilde{h}^{\alpha\beta}$ for the manipulation while we
prefer using $h^{\alpha\beta}$, the transformation is as follows
\begin{eqnarray}
\tilde{h}^{\alpha\beta}=h^{\alpha\beta}-\frac{1}{2}\eta^{\alpha\beta}h.
\end{eqnarray}
The corresponding first order harmonic gauge is
$\partial_{\beta}\bar{h}^{\alpha\beta}=0$. Moreover, we will
substitute the mass quadrupole moment $I_{ij}$ and current
quadrupole moment $J_{ij}$ as determined by
Poisson~\cite{Poisson}:
\begin{eqnarray}
I^{ij}=\frac{32}{45}M^{6}\dot{E}^{ij},\quad{}
J^{ij}=\frac{8}{15}M^{6}\dot{B}^{ij},
\end{eqnarray}
where $M$ is the mass of the black hole. The value of the tidal
heating is something like
$I^{ij}\dot{E}_{ij}+J^{ij}\dot{B}_{ij}\simeq{}\dot{E}^{2}_{ij}+\dot{B}^{2}_{ij}$,
where $\dot{E}^{2}_{ij}$ means $\dot{E}^{ij}\dot{E}_{ij}$,
likewise for $\dot{B}^{2}_{ij}$. While for the total time
derivative of $\partial_{0}(I^{ij}E_{ij})$ and
$\partial_{0}(J^{ij}B_{ij})$, they correspond to a change of a
state function, this kind of reversible changes do not involve
energy dissipation. Similarly, for the quantities
$\partial_{0}(I^{ij}B_{ij})$ and $\partial_{0}(J^{ij}E_{ij})$.

\section{Energy-momentum tensors of $B$ and $V$}

\subsection{Alternative gravitational energy-momentum tensor $V$}
Let $x^{\mu}=(t,x,y,z)$ and using a RNC Taylor expansion around
any point, consider all the possible combinations of the small
region in vacuum, the total energy-momentum density pseudotensor
is in general expressed as
\begin{equation}
{\cal{T}}_{\alpha}{}^{\beta}=\kappa^{-1}G_{\alpha}{}^{\beta}
+(a_{1}\tilde{B}_{\alpha}{}^{\beta}{}_{\xi\kappa}
+a_{2}\tilde{S}_{\alpha}{}^{\beta}{}_{\xi\kappa}
+a_{3}\tilde{K}_{\alpha}{}^{\beta}{}_{\xi\kappa}
+a_{4}\tilde{T}_{\alpha}{}^{\beta}{}_{\xi\kappa})x^{\xi}x^{\kappa}+{\cal{}O}(\mbox{Ricci},x)+{\cal{}O}(x^{3}),
\label{25Sep2008}
\end{equation}
where $a_{1}$ to $a_{4}$ are constants. Since our concern is the
vacuum case, so $G_{\alpha\beta}=0=T_{\alpha\beta}$. Then the
first order linear in Ricci terms ${\cal{}O}({\mbox{Ricci}},x)$
vanish. The lowest order non-vanishing term is of second order,
and compared to this in the small region limit we ignore the third
order terms ${\cal{}O}(x^{3})$.  It should be noted that
${\cal{T}}_{\alpha}{}^{\beta}$ in (\ref{24bDec2021}) or
(\ref{25Sep2008}) is a pseudotensor, but in the Taylor expansion
on the right hand side in (\ref{25Sep2008}) the coefficients of
the various powers of $x$ are tensors. As argued in~\cite{Deser},
$\partial^{2}_{\mu\nu}{\cal{T}}_{\alpha}{}^{\beta}(0)$ must be
some linear combination of 4 tensors, here we use \{$\tilde{B}$,
$\tilde{S}$, $\tilde{K}$, $\tilde{T}$\}. From now on, we only keep
the second order term and drop the others. There are two physical
conditions which can constrain the unlimited combinations between
\{$\tilde{B}$, $\tilde{S}$, $\tilde{K}$, $\tilde{T}$\}: 4-momentum
conservation and positivity, both considered in the small region
vacuum limit (i.e., not restricted to a 2-sphere)

First~condition: energy-momentum conservation. Consider
(\ref{24bDec2021}) and (\ref{25Sep2008}) in vacuum
\begin{eqnarray}
0=\partial_{\beta}\,\mathbf{t}_{\alpha}{}^{\beta}
=\frac{1}{4}(a_{1}-2a_{2}+3a_{3}-a_{4})g_{\alpha\beta}x^{\beta}R^{2}_{\lambda\sigma\rho\tau}.\label{24eFeb2010}
\end{eqnarray}
Therefore, the constraint for the conservation of the
energy-momentum density is
\begin{equation}
a_{4}=a_{1}-2a_{2}+3a_{3}.\label{9Jan2009}
\end{equation}
No single element from
$\{\tilde{B},\tilde{S},\tilde{K},\tilde{T}\}$ can satisfy
(\ref{24eFeb2010}), however certain linear combinations of them
can. Eliminate $\tilde{T}$ which is absorbed by $\tilde{B}$,
$\tilde{S}$ or $\tilde{K}$, comparing (\ref{24bDec2021}) and using
(\ref{9Jan2009}) rewrite (\ref{25Sep2008})
\begin{eqnarray}
\mathbf{t}_{\alpha\beta}
&=&\left[a_{1}(\tilde{B}_{\alpha\beta\xi\kappa}
+\tilde{T}_{\alpha\beta\xi\kappa})
+a_{2}(\tilde{S}_{\alpha\beta\xi\kappa}
-2\tilde{T}_{\alpha\beta\xi\kappa})
+a_{3}(\tilde{K}_{\alpha\beta\xi\kappa}
+3\tilde{T}_{\alpha\beta\xi\kappa})\right]x^{\xi}x^{\kappa}\nonumber\\
&=&(a_{1}B_{\alpha\beta\xi\kappa} +a_{2}S_{\alpha\beta\xi\kappa}
+a_{3}K_{\alpha\beta\xi\kappa})x^{\xi}x^{\kappa}\nonumber\\
&=&\left[(a_{1}+a_{3})B_{\alpha\beta\xi\kappa}
+(a_{2}-a_{3})S_{\alpha\beta\xi\kappa}
+a_{3}W_{\alpha\beta\xi\kappa})\right]x^{\xi}x^{\kappa}.\label{25aSep2008}
\end{eqnarray}
Consider all the possible expressions for the pseudotensors (some
of which explicitly included the flat metric), there indeed does
appear linear combinations of these three
tensors~\cite{SoNesterPRD2009, MTW, SoCQG2009}.  Explicitly one
can use either $\{B,S,K\}$ or $\{B,S,W\}$. We prefer the latter
because one can even define a new 4th rank energy-momentum tensor
as follows
\begin{eqnarray}
T_{\alpha\beta\mu\nu}:=B_{\alpha\beta\mu\nu}+\alpha{}W_{\alpha\beta\mu\nu},
\end{eqnarray}
where $\alpha$ is a constant. When $\alpha=1$. then $V$ is
recovered.

Second~condition: non-negative gravitational energy. For
simplicity, we use a small sphere. For any quantity at $t=t_{0}$
we consider the limiting value for the radius
$r:=\sqrt{x^{2}+y^{2}+z^{2}}$. The 4-momentum at time $t=0$ is
\begin{eqnarray}
2\kappa{}P_{\mu}=\int{}\mathbf{t}^{\rho}{}_{\mu\xi\kappa}x^{\xi}x^{\kappa}d\Sigma_{\rho}
=\mathbf{t}^{0}{}_{\mu{}ij}\int{}x^{i}x^{j}d^{3}x
=\mathbf{t}^{0}{}_{\mu{}ij}\delta^{ij}\,\frac{4\pi{}r^{5}}{15},\label{5Jan2009}
\end{eqnarray}
Thus, from (\ref{25aSep2008})
\begin{equation}
P_{\mu}=(-E,\vec{P})=-\frac{r^{5}}{60G}\left[
(a_{1}+a_{3})B_{\mu{}0ij}+(a_{2}-a_{3})S_{\mu{}0ij}\right]\delta^{ij},\label{25dSep2008}
\end{equation}
The energy-momentum values associated with $\{B,S,W\}$ are
\begin{eqnarray}
B_{\mu{}0ij}\delta^{ij}=(E^{2}_{ab}+H^{2}_{ab},2\epsilon_{cab}E^{ad}H^{b}{}_{d}),
{}S_{\mu{}0ij}\delta^{ij}=-10(E^{2}_{ab}-H^{2}_{ab},0),
{}W_{\mu{}0ij}\delta^{ij}=0.\label{25bSep2008}
\end{eqnarray}
Here we emphasize that in a small sphere region, the
energy-momentum of $B$ or $V$ is inside the light-cone,
$-P_{0}\geq|\vec{P}|\geq0$. Observing (\ref{25dSep2008}),
basically we are considering the positive energy, $B$ and $V$
already satisfy this condition and the remaining job is to find
$\{a_{2},a_{3}\}$. Equation (\ref{25bSep2008}) shows that
$S_{\mu{}0ij}\delta^{ij}$ cannot ensure positivity, since we
should allow for any magnitude of $|E_{ab}|$ and $|H_{ab}|$. The
only possibility for (\ref{25dSep2008}) to guarantee positivity is
to require $a_{1}+a_{3}\geq{}10|a_{2}-a_{3}|\geq0$. However, if we
insist the pure ``Bel-Robinson  momentum"~\cite{Szabados},
obviously, we only have one choice $a_{2}=a_{3}$.

\subsection{Computing energy-momentum and angular momentum}
The Papapetrou pseudotensors~\cite{SoCQG2009} gives a certain
linear combination of $(B,V)$ or $(B,W)$:
$2\kappa{}P^{\alpha\beta}=\frac{1}{9}(4B^{\alpha\beta}{}_{\xi\kappa}
-V^{\alpha\beta}{}_{\xi\kappa})x^{\xi}x^{\kappa}$. The energy
using (\ref{5Jan2009}) in a small sphere is
\begin{eqnarray}
P_{0}&=&-\frac{r^{5}}{540G}(4B_{00ij}-V_{00ij})\delta^{ij}\nonumber\\
&=&-\frac{r^{5}}{540G}(3B_{00ij}-W_{00ij})\delta^{ij}\nonumber\\
&=&-\frac{r^{5}}{180G}B_{00ij}\delta^{ij},
\end{eqnarray}
where
$(B_{00ij}-V_{00ij})\delta^{ij}\equiv0\equiv{}W_{00ij}\delta^{ij}$.
Before we proceed, one might question that perhaps $V$ is
superfluous since $B$ and $V$ have so far shown no distinction. We
claim that $B$ and $V$ are distinct because they are constructed
from different basic quadratic curvatures
$(\tilde{B},\tilde{S},\tilde{K},\tilde{T})$:
$B=\tilde{B}+\tilde{T}$ and $V=\tilde{S}+\tilde{K}+\tilde{T}$.
Strictly speaking, we claim $B$ and $V$ are fundamentally
different~\cite{SoCQG2009}. But this raises a question regarding
how to see the distinction clearly. we realize that it is
impossible to distinguish $B$ and $V$ if we consider 4-momentum or
angular momentum in a small sphere. So we change our strategy to
evaluating these physical quantities in other quasilocal volume
elements (see Table 1).

We claim $B$ and $V$ can have different energy values, for
instance, in a small box with different dimensions. Here we give a
concrete example: let $a=b$, $c=a+\Delta$ and $|\Delta|<<a$. The
energy for substituting $\mathbf{t}$ by $B$ is
$P^{B}_{0}\simeq\frac{a^{5}}{12}(B^{0}{}_{0ij}\delta^{ij}+\frac{2\Delta}{a}B^{0}{}_{033})$.
Similarly for $V$,
$P^{V}_{0}\simeq\frac{a^{5}}{12}(V^{0}{}_{0ij}\delta^{ij}+\frac{2\Delta}{a}V^{0}{}_{033})$.
Thus, generally, $B$ and $V$ are separable:
$P^{V}_{0}-P^{B}_{0}\simeq\frac{a^{4}\Delta}{6}W^{0}{}_{033}\neq0$.
Following the restriction that the quasilocal energy-momentum must
be a multiple of ``Bel-Robinson momentum"~\cite{Szabados}. We can
fulfill this requirement using either $B$ or $V$ in a small region
for a perfect sphere or a box with $a\equiv{}b\equiv{}c$, i.e., a
cube~\cite{Garecki}, for a cylinder or half-cylinder we need
$h\equiv\sqrt{3}a$. These are desirable results, but
unfortunately, we  lose the distinction between $B$ and $V$ again.

Is it possible to keep a multiple of ``Bel-Robinson momentum" and
still able to tell the difference between $B$ and $V$ naturally?
Yes, it is possible: we turn to examining the angular momentum
(see e.g., \S20.3 in~\cite{MTW}) which can be defined as follows
\begin{eqnarray}
J^{\mu\nu}:=\int(x^{\mu}\mathbf{t}^{\nu0}{}_{\xi\kappa}
-x^{\nu}\mathbf{t}^{\mu0}{}_{\xi\kappa})x^{\xi}x^{\kappa}d^{3}x.
\end{eqnarray}
where $\mathbf{t}$ can be $B$ or $V$. According to Table 1, we
observe that the angular momentum vanishes for a perfect sphere,
ellipsoid, box or cylinder. Conversely, both hemi-sphere and half
cylinder $(h\equiv\sqrt{3}a)$ have non-vanishing angular momentum.
In these regions, the angular momentum values for $B$ and $V$ are
distinguishable, i.e., $V$ is no longer superfluous. Moreover, we
remark that for a hemi-sphere, if we substitute $\mathbf{t}$ by
the completely symmetric $B$,
$J^{12}_{B}=\frac{\pi}{12}(B_{0123}-B_{0213})r^{6}\equiv{}0$.
However, if consider $V$,
$J^{12}_{V}=\frac{\pi}{12}(V_{0123}-V_{0213})r^{6}\neq0$
generally. Thus, the difference between $B$ and $V$ becomes
sharply manifest, showing that in this case $V$ is essential, not
redundant.

\begin{table}
\begin{tabular}{ll}
\hline Perfect-
&$P_{\mu}=\frac{4\pi}{15}t^{0}{}_{\mu{}ij}\delta^{ij}a^{5}$,\quad{}$r\in[0,a],~\theta\in[0,\pi],~\phi\in[0,2\pi]$\\
sphere & $J^{0m}=(0,0,0)$,\quad{} $(J^{12},J^{13},J^{23})=(0,0,0)$ \\

\hline Ellipsoid
&$P_{\mu}=\frac{4\pi}{15}(t^{0}{}_{\mu{}11}a^{2}+t^{0}{}_{\mu{}22}b^{2}+t^{0}{}_{\mu{}33}c^{2})abc$,\quad
$x\in[-a,a],~y\in[-b,b],~z\in[-c,c]$\\
& $J^{0m}=(0,0,0)$,\quad{} $(J^{12},J^{13},J^{23})=(0,0,0)$ \\

\hline Hemi-
&$P_{\mu}=\frac{2\pi}{15}t^{0}{}_{\mu{}ij}\delta^{ij}a^{5}$,\quad{}$r\in[0,a],~\theta\in[0,\pi/2],~\phi\in[0,2\pi]$\\
sphere & $J^{0m}=\frac{\pi}{24}(2t^{0}{}_{013},2t^{0}{}_{023},t^{0}{}_{0ij}\delta^{ij}+t^{0}{}_{033})r^{6}$ \\
& $J^{12}=\frac{\pi}{12}(t^{1}{}_{023}-t^{2}{}_{013})r^{6}$,
$J^{13}=\frac{\pi}{24}(t^{1}{}_{0ij}\delta^{ij}+t^{1}{}_{033}-2t^{3}{}_{013})r^{6}$, \\
&$J^{23}=\frac{\pi}{24}(t^{2}{}_{0ij}\delta^{ij}+t^{2}{}_{033}-2t^{3}{}_{023})r^{6}$ \\

\hline Box
&$P_{\mu}=\frac{1}{12}(t^{0}{}_{\mu11}a^{2}+t^{0}{}_{\mu22}b^{2}+t^{0}{}_{\mu33}c^{2})abc$
,~$x\in[-\frac{a}{2},\frac{a}{2}],~y\in[-\frac{b}{2},\frac{b}{2}],~z\in[-\frac{c}{2},\frac{c}{2}]$\\
&$J^{0m}=(0,0,0)$,\quad{} $(J^{12},J^{13},J^{23})=(0,0,0)$ \\

\hline Cylinder
&$P_{\mu}=\frac{\pi}{4}t^{0}{}_{\mu{}ij}\delta^{ij}a^{4}h+\frac{\pi}{12}t^{0}{}_{\mu33}(h^{2}-3a^{2})a^{2}h$,
~~$\rho\in[0,a],~\varphi\in[0,2\pi],~z\in[-\frac{h}{2},\frac{h}{2}]$\\
&$J^{0m}=(0,0,0)$,\quad{} $(J^{12},J^{13},J^{23})=(0,0,0)$ \\

\hline Half-
&$P_{\mu}=\frac{\pi}{8}t^{0}{}_{\mu{}ij}\delta^{ij}a^{4}h
+\frac{\pi}{24}t^{0}{}_{\mu33}(h^{2}-3a^{2})a^{2}h$,~~$\rho\in[0,a],
~\varphi\in[0,\pi],~z\in[-\frac{h}{2},\frac{h}{2}]$ \\
cylinder &$J^{01}=\frac{4}{15}t^{0}{}_{012}a^{5}h$,
$J^{02}=\frac{1}{18}t^{0}{}_{033}a^{3}h^{3}+\frac{2}{15}(t^{0}{}_{011}+2t^{0}{}_{022})a^{5}h$,
$J^{03}=\frac{1}{9}t^{0}{}_{023}a^{3}h^{3}$\\
&$J^{12}=\frac{1}{18}t^{1}{}_{033}a^{3}h^{3}
+\frac{2}{15}(t^{1}{}_{011}+2t^{1}{}_{022}-2t^{2}{}_{012})a^{5}h$,
$J^{13}=\frac{1}{9}t^{1}{}_{023}a^{3}h^{3}-\frac{4}{15}t^{3}{}_{012}a^{5}h$ \\
&$J^{23}=\frac{1}{18}(2t^{2}{}_{023}-t^{3}{}_{033})a^{3}h^{3}
-\frac{2}{15}(t^{3}{}_{011}+2t^{3}{}_{022})a^{5}h$\\
\hline
\end{tabular}
\centering \caption{Energy-momentum and angular momentum in
different small regions}\label{13April2012}
\end{table}

\subsection{Counting the independent components of $\{B,V,W\}$}
Basically $B$, $V$ and $W$, are fourth rank tensor and could have
256 components. However, by symmetry, they only have a relatively
small number of independent components. The counting of the number
of independent components of $B$ has already been done, here we
claim there is no common term between $B$ and $W$, i.e.,
$\{B\}\bigcap\,\{W\}=\{\emptyset\}$. We verify this statement as
follows:

First, we count the components of $B$. In principle, $B$ is fully
symmetric, by explicit examination it reduces to 35 components.
There is a formula that directly gives this number. A $k$th rank
totally symmetric tensor in $n$ dimensional space has
$C^{n+k-1}_{k}$ components. For our case $C^{4+4-1}_{4}=35$. Since
$B$ is completely tracefreeness, there are 10 additional
constraints which reduce the number of components. Therefore, we
have left only 25  components for $B$ (another argument
see~\cite{Lobo}).

Next, we count the number of independent components of $V$. $V$
does not have the totally symmetric property, but as mentioned in
(\ref{3eDec2011}) that
$V_{\alpha\beta\mu\nu}\equiv{}V_{(\alpha\beta)(\mu\nu)}\equiv{}V_{\mu\nu\alpha\beta}$.
This reduces $V$ to 55 components. However, the completely
traceless condition gives two extra constraints indicated in
(\ref{3eDec2011}) again:
$V^{\alpha}{}_{\alpha\mu\nu}\equiv{}0\equiv{}V^{\alpha}{}_{\mu\alpha\nu}$.
Consequently we have $55-10-10=35$ for $V$.

Finally, we count the number of independent components of $W$.
Observing that $V$ and $W$ are similar. Referring to
(\ref{3eDec2011}), there should thus be at most 35 components.
However, take care an extra constraint
$W_{\alpha(\beta\mu\nu)}\equiv{}0$ which gives 25 more
constraints.  Hence we find $35-25=10$ for $W$.

\subsection{Physical meaning of the completely traceless property}
It is easy to check that $B$ and $V$ are fully trace free. We are
going to verify that this mathematical property and the physical
conservation laws are in a 1-1 correspondence in the quasilocal
limit. Consider a linear combination between
$\{\tilde{B},\tilde{S},\tilde{K},T\}$, let
\begin{equation}
A:=a_{1}\tilde{B}+a_{2}\tilde{S}+a_{3}\tilde{K}+a_{4}\tilde{T}.
\end{equation}
We observe that there are only two distinct trace because of the
symmetry:
\begin{eqnarray}
8A^{\alpha}{}_{\mu\alpha\nu}\equiv(a_{1}-2a_{2}+3a_{3}-a_{4})g_{\mu\nu}R^{2}_{\lambda\sigma\rho\tau},\quad{}
2A^{\alpha}{}_{\alpha\mu\nu}\equiv(a_{1}+a_{2}-a_{4})g_{\mu\nu}R^{2}_{\lambda\sigma\rho\tau}.
\label{24aFeb2010}
\end{eqnarray}
The totally traceless condition requires that the above two
equations vanish simultaneously
\begin{eqnarray}
0=a_{1}-2a_{2}+3a_{3}-a_{4},\quad{}
0=a_{1}+a_{2}-a_{4}.\label{24cFeb2010}
\end{eqnarray}
The first equation in (\ref{24cFeb2010}) is the same as
(\ref{9Jan2009}), which indicates one of the mathematical
traceless conditions identical to the energy-momentum conservation
criterion: solving the equations in (\ref{24cFeb2010}), we obtain
$a_{2}=a_{3}$, and this is proportional to the ``Bel-Robinson
momentum" requirement found from (\ref{25dSep2008}); we have noted
that the fully tracefreeness property is related to some physical
conditions.

\subsection{The tidal heating, rate of change of momentum and spin angular momentum flux for $B$ and $V$}

The angular momentum for a perfect sphere is vanishing. However, a
slightly time dependent deformed sphere such as an ellipsoid would
be no longer zero. In order to compute the spin angular momentum
flux, the shape will be changed from a perfect sphere to an
elliptical. Practically, the lowest order is the the quadruple
moment~\cite{Zhang}. Based on the Bel-Robinson tensor, the tidal
heating can be calculated for the Bel-Robinson tensor as follows
\begin{eqnarray}
\dot{W}_{B}&=&\oint_{\partial{}V}B^{0}{}_{fmn}\,x^{f}x^{m}x^{n}\,r\,d\Omega\nonumber\\
&=&2\oint_{\partial{}V}(\partial_{\sigma}\Gamma^{0}{}_{f\lambda}-\partial_{f}\Gamma^{0}{}_{\lambda\sigma})
(\partial_{n}\Gamma^{\lambda\sigma}{}_{f}-\partial^{\sigma}\Gamma^{\lambda}{}_{fn})\,x^{f}x^{m}x^{n}\,r\,d\Omega
\nonumber\\
&=&\oint_{\partial{}V}\left[-\frac{27}{r^{4}}I_{c\,d}\dot{E}_{p\,q}\,x^{c}x^{d}x^{p}x^{q}+10I_{ij}\dot{E}^{ij}\right]d\Omega\nonumber\\
&&+\oint_{\partial{}V}\left[-\frac{68}{3r^{4}}J_{c\,d}\dot{B}_{p\,q}\,x^{c}x^{d}x^{p}x^{q}
+\frac{32}{3r^{2}}J^{p}{}_{c}\dot{E}_{p\,d}\,x^{c}x^{d}+8J_{ij}\dot{B}^{ij}\right]d\Omega\nonumber\\
&=&\frac{2}{G}\left(I_{ij}\dot{E}^{ij}+\frac{4}{3}J_{ij}\dot{B}^{ij}\right)\nonumber\\
&=&\frac{64}{45G}M^{6}(\dot{E}^{2}_{ij}+\dot{B}^{2}_{ij}).
\end{eqnarray}
Here come to the laws of motion and precession. The rate of change
of momentum
\begin{eqnarray}
\dot{P}^{i}_{B}&=&\oint_{\partial{}V}B^{i}{}_{fmn}\,x^{f}x^{m}x^{n}\,r\,d\Omega\nonumber\\
&=&\oint_{\partial{}V}\left(2R^{i}{}_{\lambda\,m\,\sigma}R^{\lambda}{}_{f}{}^{\sigma}{}_{n}\,x^{f}x^{m}x^{n}\,r
-\frac{1}{8}R^{2}_{\lambda\sigma\rho\tau}x^{i}r^{3}\right)d\Omega\nonumber\\
&=&2\oint_{\partial{}V}(\partial_{\sigma}\Gamma^{i}{}_{m\lambda}-\partial_{m}\Gamma^{i}{}_{\lambda\sigma})
(\partial_{n}\Gamma^{\lambda\sigma}{}_{f}-\partial^{\sigma}\Gamma^{\lambda}{}_{fn})\,x^{f}x^{m}x^{n}\,d\Omega\nonumber\\
&=&\frac{2}{3}\epsilon^{i}_{a\,b}\left(I^{a\,c}\dot{B}^{b}{}_{c}-\frac{4}{3}J^{a\,c}\dot{E}^{b}{}_{c}\right)\nonumber\\
&=&\frac{128}{135G}M^{6}\epsilon^{i}{}_{a\,b}\dot{E}^{a\,c}\dot{B}^{b}{}_{c}.
\end{eqnarray}
Note that the maximum value for
$\dot{P}^{i}_{B}=2\times(256/225)M^{6}\epsilon^{i}{}_{a\,b}\dot{E}^{a\,c}\dot{B}^{b}{}_{c}$
that can satisfy inside the light-cone requirement, i.e.,
$\dot{W}_{B}\geq|\dot{P}^{i}_{B}|\geq0$. Consider the spin angular
momentum flux for the Bel-Robinson tensor
\begin{eqnarray}
\dot{J}^{i}_{B}
&=&\oint_{\partial{}V}\epsilon^{i}{}_{j\,k}\,B^{k}{}_{fmn}\,x^{j}x^{f}x^{m}x^{n}\,r\,d\Omega\nonumber\\
&=&2\oint_{\partial{}V}\epsilon^{i}{}_{j\,k}\,
R^{k}{}_{\lambda\,m\,\sigma}R^{\lambda}{}_{f}{}^{\sigma}{}_{n}\,x^{j}x^{f}x^{m}x^{n}\,r\,d\Omega\nonumber\\
&=&-\frac{4}{G}\epsilon^{i}{}_{a\,b}\left(I^{a\,c}E^{b}{}_{c}+\frac{4}{3}J^{a\,c}B^{b}{}_{c}\right).
\end{eqnarray}

Moreover, we also demonstrate these three physical quantities
$\dot{W}$, $\dot{P}^{i}$ and $\dot{J}^{i}$ for replacing the
Bel-Robinson tensor by tensors $S$ and $W$. We found that all of
them give null result. As these three quantities are real physical
phenomena, we have to acknowledge that the super-energy
Bel-Robinson tensor $B$ and tensor $V$ ($V=B+W$) contribute the
same values.

\section{Conclusion}
For describing positivity, the Bel-Robinson tensor is the best,
and perhaps has been thought to be the only possibility. We
recently proposed an alternative $V$ in such a way that it shares
the same energy-momentum as $B$ does in the small sphere limit.
One might think that $B$ and $V$ cannot be distinguished, but we
claim they can. After examining the energy found from other
2-surface such as in ellipsoid, box, cylinder and half-cylinder
$(h\neq\sqrt{3}a)$, we demonstrate that $V$ is not redundant
because $B$ and $V$ are distinguishable. However, if we insist to
achieve a multiple of pure ``Bel-Robinson momentum" from
Szabados's argument in Living Review, the distinction between $B$
and $V$ will be lost once more. For a shape such that both $B$ and
$V$ give a multiple of the pure ``Bel-Robinson momentum" we can
turn to investigate the angular momentum. Thus when replacing
$\mathbf{t}$ by either $B$ or $V$, indeed they do lead to
different angular momentum values for a hemi-sphere or
half-cylinder with $h=\sqrt{3}a$. Moreover, we emphasize that some
of the components of the angular momentum give a null result for
$B$ and a non-vanishing result for $V$. The reason is based on the
elegant completely symmetric property of $B$, while $V$ is not
fully symmetric. Thus $V$ can play an essential irreplaceable
role.

The tensors $B$ and $V$ are constructed from different fundamental
quadratic curvatures $\{\bar{B},\tilde{S},\tilde{K},\tilde{T}\}$.
As a double check, we counted the independent components of $B$
and $V$ and find that they are not the same. Finally, we discover
the necessary and sufficient conditions for $B$ and $V$: fully
tracefreeness and conservation of future pointing non-spacelike
pure ``Bel-Robinson momentum" in the small region limit.

Furthermore, we also demonstrate the tidal heating, rate of change of momentum and spin
angular momentum flux by using tensors $B$ and $V$. Once again,
they have the same results.

\section*{Acknowledgment}
The author would like to thank Dr. Peter Dobson, Professor
Emeritus, HKUST, for reading the manuscript and providing some
helpful comments. This work was supported by NSC
95-2811-M-032-008, NSC 96-2811-M-032-001, NSC 97-2811-M-032-007
and NSC 98-2811-M-008-078.

\end{document}